\documentclass[preprint,pra,letterpaper,superscriptaddress,floatfix]{revtex4}
\usepackage{graphicx,psfrag,bbm,latexsym,color,dcolumn,bm,dsfont,bbm,color,mathrsfs,bbold,latexsym,amsmath,amsfonts,
amssymb,epsfig,natbib}

\newcommand{\beq}{\begin{equation}}
\newcommand{\eeq}{\end{equation}}

\newcommand{\beqa}{\begin{eqnarray}}
\newcommand{\eeqa}{\end{eqnarray}}

\begin{document}

\title{Nonperturbative Renormalization of the Heisenberg Spin-$1/2$ Antiferromagnet on the Square Lattice}
\author{P.R. Crompton}
\affiliation{Department of Applied Maths, School of Mathematics, Univerity of Leeds, Leeds, LS2 9JT, UK.}
\vspace{0.2in}
\date{\today}

\begin{abstract}
\vspace{0.2in}
{We investigate the critical scaling of the spin-1/2 antiferromagnet on the square lattice in the easy-plane (XXZ) regime, via numerical measurements of the entanglement entropy constructed from the zeroes of a polynomial ring. We relate these results to conformal field theory predictions for the area law scaling of entanglement entropyin the vicinity of quantum critical points in gapless conformally invariant 2d quantum systems, and in gapless systems with finite Fermi surfaces. Our measurements are focused in three low-temperature regions at fixed XXZ anisotropies of $\Delta=1.01$, $\Delta=1.78$ and $\Delta=2.0$ which probe the quantum regime between the N\'eel point and Ising phase. 
}
\vspace{0.1in}
\end{abstract}

\maketitle

\section{introduction}

The spin-$1/2$ Heisenberg antiferromagnet on the square lattice has had a long association with high-temperature superconductivity \cite{and}\cite{afm2}, but there has been almost an equally long-standing controversy surrounding the topological nature of certain of its ground states \cite{p1029_1}-\cite{p7215_1}. This question has become more pressing recently due to the discovery, via STM and neutron scattering data, that spin impurities in the underdoped cuprates apparently induce a direct second order transition between the N\'eel ordered state and a condensed Bose phase \cite{exp1}\cite{exp2}, fuelling further interest in mechanisms of the quantum criticality of edge states \cite{132}\cite{133}. In contrast, the behaviour of the one dimensional analogue of this system is largely settled. Many quantum spin chain models can be solved exactly, and there is a well-known direct mapping between chains at criticality and the $SU(2)$ WZW model \cite{chain}, which leads to a distinction between gapped and gapless groundstates for integer and half-integer spin chains. What makes the 2d system inherently more involved is that whilst the Hilbert space dimension is an extensive quantity the Hopf invariant is also zero \cite{p937_1}, which leads to the emergence of two competing spin wave theories for the ground state (or more properly a chiral wave theory coupled via a Chern-Simons term in the continuum limit \cite{10}). Since the hard-core bosons of the Jordan-Wigner approach will then necessarily carry fermionic statistics in the long-distance (IR) regime of this system (in addition to the short-distance (UV) regime  \cite{142}\cite{147}) this has motivated various ideas of Bose condensation via skyrmions for the ground state of this system in the quantum regime \cite{200}\cite{p7215_1}\cite{149}. 

In this article we consider the anistropic lattice model defined via,

\beq
H = J \, \left( \,\, \sum_{i,\,j=1}^{L}{\bm S}^{\,{\bm r}}_{i} . {\bm S}^{ \, {\bm r}}_{j} + (\Delta -1){\bm S}^{z}_{i} .{\bm S}^{z}_{j} \,\, \right)
\eeq

where $i$ and $j$ are spin sites indices, $J$ is the nearest-neighbour spin interaction coupling, $\Delta$ is the XXZ spin anisotropy parameter, and ${\bm S}$ is a spin-$1/2$ operator represented by usual the Pauli matrices with spin components $ \bm{r} = (x,y,z) $. At $\Delta=1$ the zero temperature ground state is N\'eel ordered \cite{and}. By considering the convergence of the support of spin correlators as a function of $L$ it has been identified in \cite{11}\cite{138} that there is also long-range order (LRO) for anisotropy parameter values of $0<\Delta <0.13$, and $\Delta > 1.78$ (in the easy-axis and easy-plane limits). At $\Delta \sim 0$ a second order transition is expected as a function of temperature, but this is then expected to become first order when quantum fluctuations suppress this LRO \cite{13}, which has been confirmed by numerics in \cite{8}\cite{14}. 
Attempts to treat quantum fluctuations in the spin-$1/2$ Heisenberg antiferromagnet on the square lattice in the vicinity of $\Delta=1$ have focused on treating the N\'eel order parameter as the basis of a low-energy effective field theory, via the $(2+1)$-dimensional $\sigma$ model \cite{afm2}\cite{p1029_1}. In \cite{17}\cite{GetPDF5} a fugacity expansion on the dual is used to identify that a dangerously irrelevant monopole contribution will lead to a second order (quantum) transition from the N\'eel to the Valence Bond Solid phases \cite{and}. 

The focus of this article is universality of the transitions in the quantum regime of the spin-$1/2$ Heisenberg antiferromagnet on the square lattice. However, two factors make the direct numerical verification of the quantum transitions in \cite{17}\cite{GetPDF5} difficult. Firstly, it is unclear what the relevant order parameters should be \cite{133}, secondly, although lattices serve as effective short-range regulators, lattice IR cutoffs arise from statistical averaging, which makes the lattice Lorentz symmetry (of gauge theories) approximate \cite{wilson}. Although there are no explicit gauge fields involved the construction of the spin-$1/2$ Heisenberg antiferromagnet on the square lattice in (1), because the Hopf invariant is zero and the Hilbert space dimension is extensive \cite{p937_1}, it is important that an exact $U(1)$ Wick rotation exists between Euclidean and Imaginary time at the boundary of the system. This symmetry is described as an "emergent" gauge field when the system is both Lorentz and scale invariant \cite{17}\cite{GetPDF5}. Therefore, because the Lorentz symmetry is approximate on the lattice, it is more difficult to construct numerics which have the correct Hopf invariant. For example, it is not possible to guarantee that a global minimum can be reached in quantum spin systems using DMRG via the standard approach of minimising the following expectation \cite{c},

\begin{equation}
\left\langle \psi | H | \psi \right\rangle: \, | \psi \rangle \in (\mathbb{C}^{2})^{\otimes L^{2}}
\end{equation} 

Quantifying the scaling behaviour of entanglement entropy is one means of directly establishing the universality of quantum transitions (at least for quantum chains). This is because scale invariance also implies conformal invariance in $(1+1)$ dimensions. The quantum transition of spin-$1/2$ chains is characterised universally by the central charge $c$, where $c=1$ for free bosons and $c=1/2$ for free fermions. This has been verified by a number of numerical studies \cite{st3-dm}\cite{st1-dm}\cite{dm2} (although similar problems arise with ensuring the Lorentz invariance of numerical lattice methods \cite{c}). This universal scaling is quantified via the Von Neumann entropy \cite{wil}, 

\begin{equation}
S = \frac{(c + \overline{c})\chi}{6} \log \left( \frac{\xi}{a} \right) + \mathcal{O}(e^{-L/\xi})
\end{equation}
 
where the Von Neumann entropy $S$ describes the bipartite entanglement between two subsystems, $c$ and $\overline{c}$ are the holomorphic and antiholomorphic central charges of the corresponding conformal field theory at the boundary between the subsystems, $\xi$ is the correlation length (of the finite temperature system), $a$ is the lattice UV cutoff, and $\chi$ is the Euler characteristic which categorises the topology of the boundary between subsystems \cite{122}\cite{123}\cite{euler}. 
Unfortunately, there is no generalisation of the above $c$-theorem for quantum chains \cite{Z} to higher dimensions, since the Witt algebra associated with the primary (chiral) fields (of which the central extension is the Virasoro algebra) does not map directly onto the conformal anomaly in higher dimensions. The above area law scaling is therefore not universal in higher dimensions. On dimensional grounds, the coefficient in the area law scaling for the spin-$1/2$ Heisenberg antiferromagnet on the square lattice goes as $a^{1-d}$ and is not universal, but there should also be a universal term area law for the 2d system term proportional to $\xi^{1-d}$ \cite{119}\cite{122}\cite{st2}.

Since we are faced with approximate Lorentz symmetries in standard numerical lattice approaches \cite{wilson}, and the non-universal area law scaling of the entanglement entropy of the spin-$1/2$ Heisenberg antiferromagnet on the square lattice \cite{119}\cite{122}\cite{st2}, we take a new approach to identifying the quantum transitions of this system (which we have also recently applied to the quantum spin chain in \cite{me7}). For this, we construct an exact polynomial expansion for the partition function from Wick rotating the transfer matrix elements of a standard nonperturbative (Quantum Monte Carlo) ensemble. This enables us to construct two exact nonorthogonal polynomial representations of the primary chiral field separated via a nonperturbative Chern-Simons term (which corresponds to non-analytic portion of support of this polynomial) \cite{p937_1}, by identifying the quotient of this polynomial expansion. Taking the infinite volume limit of the support of this formalism corresponds to a $\zeta$-function renormalization of the partition function, as we show, and the (central) extension of the polynomial can be determined from this limit. In this article, we generate standard ensembles at three points in the quantum regime of the anisotropic spin-$1/2$ Heisenberg antiferromagnet on the square lattice ($\Delta=1.01$, $\Delta=1.78$ and $\Delta=2$ at $\beta=100$), and determine the scaling of the Von Neumann entropy (identified from the density of the polynomial zeroes) associated with the mapping on to its $(1+1)$ primary chiral field at criticality.

\section{Partition Function Zeroes of Quantum Critical Points}

In order to understand the homology of the treatment of the spin-$1/2$ antiferromagnet on the square lattice that is presented in \cite{p937_1} it is instructive to write the lattice partition function in the following Wick-rotated form \cite{me1},

\beq
\mathcal{Z} = \int\!\int \mathcal{D}\theta_{1}\mathcal{D}\theta_{2}  \,\,\, {\rm{exp}}\!\left[\,\int_{0}^{\beta} d\tau H \,\, \,\right]  \equiv \int\! \int \mathcal{D}\theta_{1}\mathcal{D}\theta_{2}  \,\,\, {\rm{exp}}\!\left[\,\int_{0}^{\beta} d\tau\,\, A_{\bm{n}}\,i\phi_{1} + B_{\bm{n}}\,i\phi_{2} -V_{\bm{n}} \,\, \right]
\eeq

\beq
A_{\bm{n}} \equiv \sum_{s,\,s',\,s''}^{T \otimes \Theta_{1} \otimes \Theta_{2}} \,\, \sum_{\sigma \in Z_{2} } \lambda_{s,s',s'',\sigma}(\bm{n})
\frac{\langle \bm{n}\oplus \bm{1}_{s\sigma} \oplus \bm{1}_{s'\sigma} \oplus \bm{1}_{s''\sigma}| \theta_{1} \rangle}{\langle \bm{n}| \theta_{1} \rangle}\,
\eeq
\beq
B_{\bm{n}} \equiv \sum_{s,\,s',\,s''}^{T \otimes \Theta_{1} \otimes \Theta_{2} } \,\, \sum_{\sigma \in Z_{2}} \lambda_{s,s',s'',\sigma}(\bm{n})
\frac{\langle  \bm{n}\oplus \bm{1}_{s\sigma} \oplus \bm{1}_{s'\sigma} \oplus \bm{1}_{s''\sigma}| \theta_{2} \rangle}{\langle \bm{n}| \theta_{2} \rangle}\,
\eeq
\beq
V_{\bm{n}} \equiv \sum_{s,\,s',\,s''}^{T \otimes \Theta_{1} \otimes \Theta_{2} } \,\, \sum_{\sigma \in Z_{2}}\lambda_{s,s',s'',\sigma}(\bm{n}) \frac{\langle
 \bm{n}\oplus \bm{1}_{s\sigma} \oplus \bm{1}_{s'\sigma} \oplus \bm{1}_{s''\sigma}| \bm{n} \rangle} {\langle \bm{n}| \bm{n} \rangle}\,
\eeq

where $A_{\bm{n}}$, $B_{\bm{n}}$ and $V_{\bm{n}}$ are defined as the compact and noncompact portions of the spin operators in (1) determined through the nonperturbative matrix elements $\lambda(\bm{n})$ and $s$ is the Euclidean-time lattice site index. To define the indices $s'$ and $s''$ a state is taken from $|\psi \rangle \in (\mathbb{C}^{2})^{\otimes L^{2}}$ and the order of these tensor products is interchanged, which is valid on the system boundary \cite{p937_1}. Locally, however, this creates a mismatch between $\theta_{i}$ and $\phi_{i}$ ($i=1,2$), hence polynomial representations of $A_{n}$ and $B_{n}$ will not be orthogonal, in general, only in the special case where $\theta_{i} = \phi_{i}$ and the system is both Lorentz and scale invariant. 

A meaningful statistical ensemble for spin-$1/2$ antiferromagnet on the square lattice can be obtained numerically using the continuous-time Quantum Monte Carlo method \cite{qmc} by generating a Markov chain from importance sampling using the following local transfer matrices \cite{t1},

\beq
\label{transfer}
{\cal{Z}} = \prod_{i=1,j=1,t=1}^{L,L,T} \left( \begin{array}{cccc}
p_{\{i,j\},\{i+1,j\}} & p_{\{i,j\},\{i,j+1\}} \\
p_{\{i+1,j\},\{i+1,j+1\}} & p_{\{i,j+1\},\{i+1,j+1\}} \end{array} \right)
\eeq

\beq
\label{p1}
p_{\{i,j\},\{i+1,j\}} = \left( \begin{array}{cccc}
{\rm {exp}}(-\frac{\Delta\tau J_z}{2}) & 0 & 0 & 0\\
0 & {\rm {cosh}}(\frac{\Delta\tau J_{xy}}{2}) & {\rm {sinh}}(\frac{\Delta\tau J_{xy}}{2}) & 0 \\
0 & {\rm {sinh}}(\frac{\Delta\tau J_{xy}}{2}) & {\rm {cosh}}(\frac{\Delta\tau J_{xy}}{2}) & 0 \\
0 & 0 & 0 & {\rm {exp}}(-\frac{\Delta\tau J_z}{2})
\end{array} \right)
\eeq
\beq
\label{p2}
p_{\{i,j\},\{i+1,j+1\}} = \left( \begin{array}{cccc}
{\rm {exp}}(-\frac{\Delta\tau J_z}{2}) & 0 & 0 & {\rm {cosh}}(\frac{\Delta\tau J_{xy}}{2}) \\
0 & 0 & {\rm {sinh}}(\frac{\Delta\tau J_{xy}}{2}) & 0 \\
0 & {\rm {sinh}}(\frac{\Delta\tau J_{xy}}{2}) & 0 & 0 \\
{\rm {cosh}}(\frac{\Delta\tau J_{xy}}{2}) & 0 & 0 & {\rm {exp}}(-\frac{\Delta\tau J_z}{2})
\end{array} \right)
\eeq

where $p_{\{i,j\},\{i+1,j+1\}} \leftrightarrow p_{\{i,j\},\{i+1,j\}}$ and $p_{\{i+1,j\},\{i+1,j+1\}} \leftrightarrow p_{\{i,j\},\{i+1,j+1\}}$ for $J_z \leftrightarrow  J_{xy}$, $J_{z}/J_{xy} = \Delta$, and $\Delta\tau$ is the Euclidean-time lattice spacing. Discrete steps in Euclidean-time are exchanged in the continuous-time method with discrete spin flips, hence the Markov chain that is generated is ergodic in probability space (Imaginary time) but not in Euclidean-time \cite{phys}. The nonpertubative matrix elements realised in this Markov chain can be identified at each step $t$, and up to a normalising factor of $\beta J$ (the inverse temperature and bipartite lattice interaction coupling) these matrix elements are defined via, 

\beq
\label{newtransfer}
p_{\{i,j\},\{i+1,j\}}^{(t)} = \left( \begin{array}{cccc}
1 & 0 & 0 & 0\\
0 & \lambda_{t,i,j}  & 1 - \lambda_{t,i,j} & 0 \\
0 & 1 -\lambda_{t,i,j} & \lambda_{t,i,j} & 0 \\
0 & 0 & 0 & 1 \end{array} \right)
\eeq

The Wick-rotation of these elements is then defined by compactifying the Euclidean-time boundary conditions of the
lattice partition function from $(0,\beta] \rightarrow \theta = [-\pi,\pi]$, via the following definition of the trace of the transfer matrix $P$ (defined as the transfer matrix for the whole of the $L^{2}$ volume), 

\beq
{\rm Tr} P = \sum_{k=1}^{L^{2}}  ( 1 -{\lambda_k}^2 )( 1 -(\Delta\lambda_k)^2 ), \quad \mathcal{Z} = \int \mathcal{D}\theta  \,\,\, P 
\eeq

where $k$ is the lattice site index on $\Theta_{1}\otimes \Theta_{2}$, and $\Delta$ is the spin anisotropy parameter. This formalism relates the matrix elements in (11) to those in (4).
The assumption for this Wick rotation prescription  is that each local plaquette on $i\otimes j$ can be smoothly deformed into $\mathbb{C}^{2}$ via the spin anisotropy parameter $\Delta$. This is not true, since any two states in the system are not orthogonal in general. Therefore, we recover an exact expansion in $\beta J$ from these operators that is only analytic up to $\phi_{1}\otimes \phi_{2}$ (rather than $\theta_{1}\otimes \theta_{2}$). However, this allows us to probe the criticality of the primary chiral field of the nonperturbatively realised system that corresponds to this singularity. 

In the replica method \cite{122}, the Von Neumann entanglement entropy of an $n$-fold Riemann sheet (such as $\mathbb{R}^{2}$ for $n=L^{2}$) is defined by identifying the analytic continuation properties of the lattice partition function with respect to Euclidean-time, 

\begin{equation}
S = - \lim_{n\rightarrow 1} \frac{\partial}{ \partial n } \left( \frac{\mathcal{Z}_{n}}{(\mathcal{Z})^{n}}\right)
\end{equation}

where $\mathcal{Z}_{n}$ is the lattice partition function for a subsystem consisting of $n$ disjoint unions of $\mathcal{Z}$. We go slightly further than the usual replica argument for defining the entanglement entropy, in our approach, by constructing a polynomial ring for $\mathbb{C}^{2}$. This removes the assumption that the partition function has to be factorisable in the above denominator \cite{300}. This is particularly important for the spin-$1/2$ antiferromagnet on the square lattice, because general states in this system are defined by two non-orthogonal polynomials, hence, the entanglement entropy defined in the above limit is degenerate (even though the branch points are resolved in $\mathbb{R}^{2}$). The relevant quantity for determing the entanglement entropy in our formalism is the logarithm of ${\rm {det}} P$. An analogous expression to the replica limit is then obtained by analytically continuing the following function from large $s$ to $s=0$,

\beq
\sum_{k=1}^{L^{2}} \Lambda_{k}^{-s} \, {\rm{ln}}\, \Lambda_{k}
\eeq

where $\Lambda_{k} = ( 1 -{\lambda_k}^2 )( 1 -(\Delta\lambda_k)^2 )$. This amounts to the renormalization of the singularities of the support of the free energy of the partition function via a $\zeta$-function prescription, where, 

\begin{equation}
\zeta_{P}(s) = \sum_{k=1}^{L^{2}} \Lambda_{k}^{-s} = \frac{1}{\Gamma(s)} \int_{0}^{\infty} \mathcal{D}\theta \,\, {\rm{Tr}} \, (e^{-\theta P}) \, \theta^{s-1}
\end{equation}

However, although the logarithm of $\det P$ (free energy) is completely defined by the first derivative of this $\zeta$-function at $s=0$, the corresponding free energy minima is not unique since any number of the higher moments of $\zeta_{P}(s)$ can also be nonzero. Hence, to uniquely determine the entanglement entropy a further prescription must be given for resolving the choices of branch in $k$ for all of these higher moments at a different point on the fundamental strip, namely at $s=1$, 

\beq
\left. \frac{ d^{k} \zeta_{P}(s) }{ ds^{k} } \right|_{s=1}  = \int_{0}^{\infty} \mathcal{D} \theta \,\, \Lambda(\theta) \,\, {\rm {ln}}^{k}(\theta) , \quad k = 0 \,\, ... \,\, L^{2}
\eeq

This prescription is implemented by constructing a polynomial in $\theta$ consisting of $k$ terms - the zeroes of this polynomial ring therefore correspond to elements of the quotient $\mathbb{C}^{2}$ (rather than $\mathbb{R}^{2}$). Hence, the meromorphic convergence of the zeroes of this polynomial uniquely specifies the resolution of the branches in $k$, via,

\beq
\label{entropy}
{\rm{ln}}\, {\rm{det}} P  =  -\int^{\infty}_{0} \Lambda(\theta) \,\, {\rm{ln}}(\theta) \,\, d\theta \quad
 \rightarrow \quad S = - \int_{0}^{\infty} \mathcal{D} \theta\,\, \Lambda(\theta) \ln \Lambda(\theta) \,\, 
\eeq

which is the familiar Von Neumann form of the entanglement entropy. 

Practically, the support of the logarithm of $ {\rm{det}} P$ can be investigated via the scaling of the zeroes density of the eigenvalue problem ${\rm{det}}(P - (\beta J)^{2} ) =0$, since the divergences of the logarithm of $ {\rm{det}} P$ occur where this polynomial ring has its zeroes. The zeroes are obtained numerically by rootfinding the characteristic polynomial equation for this eigenvalue problem, where the characteristic polynomial coefficients are obtained from powers of ${\rm Tr} P$ by using Newton's identities \cite{me7}\cite{me55},

\beq
\label{poly}
{\cal{Z}} = \sum_{k=0}^{L^{2}} \, \langle c_{k} \rangle \, (\beta J)^{2k},  \quad k\, c_{k} + {\rm Tr} P^{k} + \sum_{n=1}^{k-1} c_{n} {\rm Tr} P^{k-n} = 0, \quad  n=1 ... L^{2}
\eeq

This polynomial zeroes approach is similar in spirit to Lee and Yang's determination of the zeroes of the Ising partition function \cite{l+y2}, although the singularities in our formulation correspond to elements of the quotient $\mathbb{C}^{2}$ (rather than $Z_{2}$). In both cases the polynomial zeroes are constrained to lie on the unit circle in the complex plane of the polynomial expansion parameter $X$. This is because the quotient of the polynomial representations of each model and the $Z_{2}$ symmetry of their spin operators form a polynomial ring, $\mathbb{C} \equiv \mathbb{R}[X]/X^{2} +1$. For the Ising model this constraint is an identity, and all the zeroes lie on the unit circle, but for our formulation this constraint only applies at criticality, since $PGL_{2}(\mathbb{C})\supseteq PSU_{2} \cong SO(3)$. Motivated by this, we apply a simple difference relation to define the zeroes density $\Lambda(\beta J)$ of our lattice systems along the unit circle \cite{lambda1},

\beq
\label{den}
\Lambda(\phi) \equiv \Lambda \left( \frac{\phi_{k+1}+\phi_{k}}{2} \right) =
\frac{1}{L^{2}(\phi_{k+1}-\phi_k)}
\eeq

where $k$ is the sequential index assigned to the zeros along the locus and $\phi$ is the angle subtended from the real-$(\beta J)^2$ axis to a given zero. The scaling of the zeroes density describes the renormalization group scale transformations of the lattice system, hence,

\beq
\label{asymp1}
\Lambda(\beta,J,\Delta,L) = L^{\tilde{c}} \Lambda(\beta L, J L, \Delta)
\eeq

where $\tilde{c}$ is the scaling exponent in the vicinity of a second order fixed point. Therefore, 
\beq
\label{asymp}
\lim_{J,\beta \rightarrow 0} \Lambda(J) = J^{\tilde{c}} ( 1 - J\, ... \,) \quad \Rightarrow \quad {\rm{ln}} \Lambda(\phi) = \tilde{c}\, / \phi + \, ...
\eeq

\section{numerical results}

We have generated lattice ensembles of the spin-$1/2$ antiferromagnet on the square lattice at several different values of length $L=\{12,16,20,24,32,40\}$, at the inverse temperature  $\beta=100$, at strong coupling (where $J=1$) and at three different fixed values of the XXZ anisotropy parameter  $\Delta=\{1.01, 1.78, 2.0\}$ using the continuous-time Quantum Monte Carlo method \cite{qmc}. We have constructed the characteristic polynomial coefficients, defined in (12), for a polynomial ring whose singularities form the quotient $\mathbb{C}^{2}$ using the numerical transfer matrix entries generated for the Markov process of each ensemble, following (18). We have applied standard rootfinding techniques to this polynomials to find its zeroes, and have constructed the simple difference relation for the zeroes density along the unit circle in the complex expansion parameter plane defined in (19). This zeroes density relation equivalently defines the Von Neumann entanglement entropy of the spin-$1/2$ antiferromagnet on the square lattice in (17), for the entanglement defined between the set of $L^{2}$ disjoint subsystems of the square lattice (that are analytic in the expansion parameter $\beta J$) and the rest of the ensemble. Each of these subsystems is therefore defined nonperturbatively by the mismatch between $\phi_{1}\otimes\phi_{2}$ and $\theta_{1}\otimes\theta_{2}$ in (4). This approach allows us to probe the entanglement entropy scaling of the primary chiral field associated with $\mathbb{C}^{2}$ directly, even though a general state of the spin-$1/2$ antiferromagnet on the square lattice system is defined by two non-orthogonal polynomials \cite{p937_1}. 

Whilst all of the zeroes we have evaluated in the complex plane automatically live on the surface traced out by $\phi_{1}\otimes\phi_{2}$ (from the definition of the polynomial in (12)) the real test of this numerical method is whether the zeroes density (entanglement entropy) scales in any meaningful way along this surface. As with the Ising model \cite{l+y2}, the zeroes polynomial in (12) defines the extension of the expansion parameter $X (=\beta J)$, and this extension becomes the central extension when the lattice ensembles we have generated are both Lorentz and scale invariant. Hence, it is only if the zeroes demonstrate some symmetry by lying on a well defined locus in the complex plane that the extension of the system can be quantified via the scaling of the zeroes density.

In Figure 1 the polynomial zeroes are plotted in the complex-$(\beta J)^2$ plane for three lattice ensembles generated at $\Delta=\{1.01, 1.78, 2.0 \}$. 
For the ensemble generated at $\Delta=1.01$ in Figure 1 all of the zeroes lie on the unit circle, hence the entanglement entropy scaling can be quantified via the zeroes density along this locus. However, whilst for $\Delta=1.78$ and $\Delta=2.0$ a subset of the zeroes lie on the unit circle, a further subset of the zeros lie on (or to the exterior of) a line at a fixed angle $\phi'$ from the real-$(\beta J)^2$ axis. There is therefore a mismatch between $\phi_{1}\otimes\phi_{2}$ and $\theta_{1}\otimes\theta_{2}$ for these ensembles. The reason for this is that in (12) the polynomial is assumed to be analytic in $\Delta$ and $J$ but it is only when the system is at criticality (Lorentz and scale invariant) that the scaling is critical in both variables. Hence, further from the N\'eel point scaling in $\Delta$ becomes subcritical and the critical point develops at $\phi =\phi'$. 

In Figure 2 we directly test the scaling hypothesis for $\phi_{1}\otimes\phi_{2}$. We rescale the pseudo-critical value $\phi_{\phi'}$ by subtracting from it the value of the zero closest the real-$(\beta J)^{2}$ axis $\phi_{0}$, and plot this difference $(\phi_{\phi'} - \phi_{0})$ against $1/L^{2}$. There should be no evidence of any meaningful scaling if the mismatch between $\phi_{1}\otimes\phi_{2}$ and $\theta_{1}\otimes\theta_{2}$ is a purely numerical artifact. However, the result in Figure 2 indicates a clear linear confining flux between these two branch points, with a string tension of $\sigma=68(4)$ for $\Delta=1.78$, and $\sigma =78(7)$ for $\Delta=2.0$, in lattice units. The results in Figure 1 and Figure 2 strongly suggest that it is meaningful to model the general states of the spin-$1/2$ antiferromagnet on the square lattice system either by two non-orthogonal polynomials \cite{p937_1}, or equivalently by a chiral wave theory coupled via a nonperturbative Chern-Simons term in the continuum limit \cite{10}, since the polynomial zeroes we have measured do have a well defined extension, which links the UV to the IR via statistical transmutation  \cite{200}\cite{p7215_1}\cite{149}.

\begin{figure}
\epsfxsize=3.5 in\centerline{\epsffile{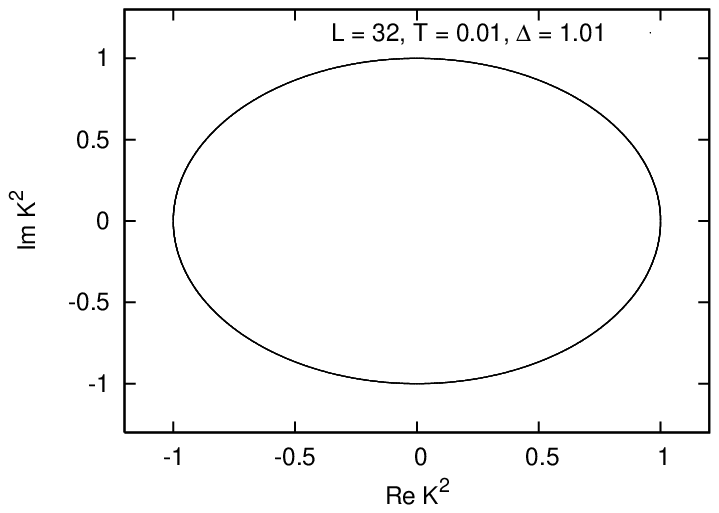}}
\epsfxsize=3.5 in\centerline{\epsffile{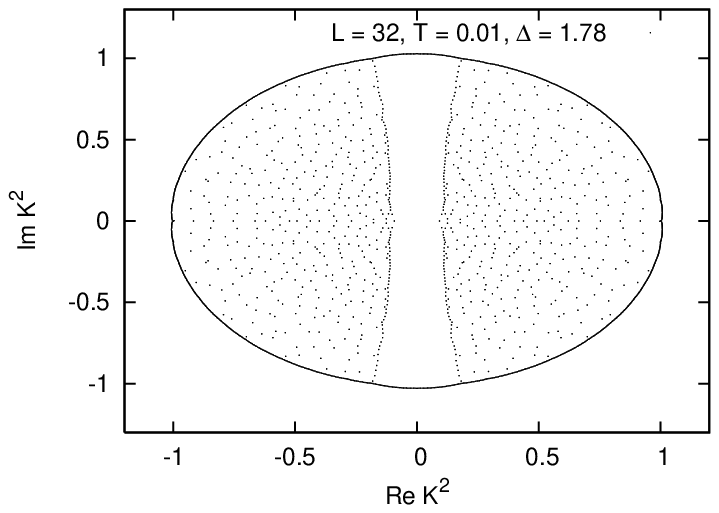}}
\epsfxsize=3.5 in\centerline{\epsffile{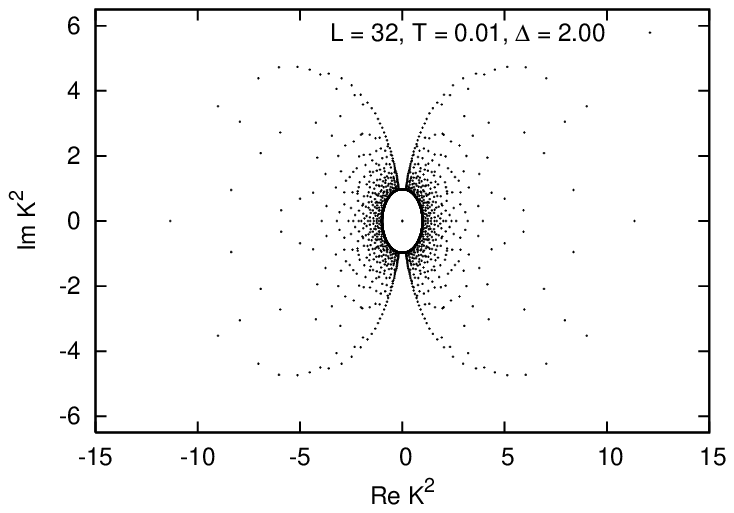}}
\caption{The polynomial zeros of the spin-1/2 antiferromagnet on the square lattice plotted in the complex-$K^{2}$ plane (where $K = \beta J$) for three different fixed values of the XXZ anisotropy parameter $\Delta=\{1.01, 1.78, 2.0\}$. The lattice size of all three systems is $L^2 = 32\times 32$, and the inverse temperature is kept fixed at $\beta=100$. A subset of the zeros lie on a locus, corresponding to the unit circle in the complex-$K^{2}$ plane, for all three ensembles. }
\end{figure}

\begin{figure}
\epsfxsize=4.5 in\centerline{\epsffile{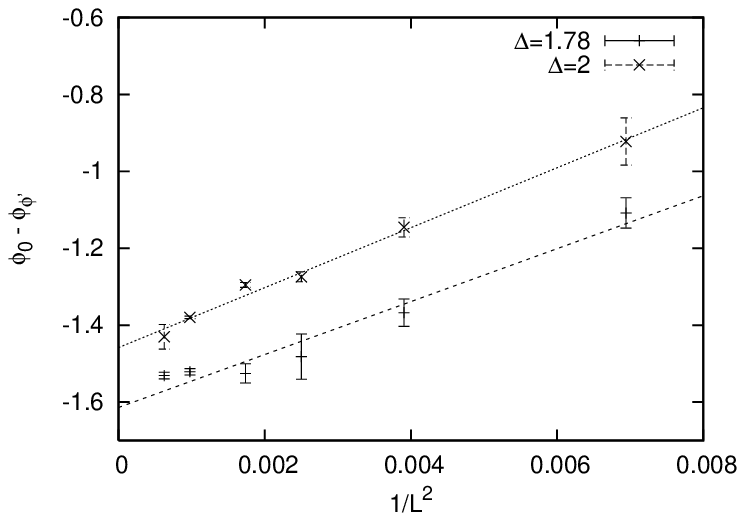}}
\caption{Pseudo-critical scaling of the difference in the critical values of the zeroes, $\phi_0$ and $\phi_{\phi'}$, as a function of the inverse lattice size squared, $1/L^2$. }
\end{figure}

On the basis of dimensional analysis \cite{122}\cite{st2}, the general picture of entanglement entropy scaling in $(2+1)$ dimensions is of a leading area law term with a prefactor of $a^{1-d}$ (where $a$ is the lattice UV cutoff), which is therefore not universal, and a subleading universal area law term propotional to $\xi^{1-d}$ (where $\xi$ is the correlation length) of the form of (3). This general picture is therefore modified if the system is either gapped or gapless \cite{118}. In \cite{euler}, conformally invariant systems at criticality have been found to satisfy this general picture. However, for the spinless gapped systems considered in \cite{112}, whilst the leading contribution is as above, the subleading behaviour differs in the quantum critical and non-critical regimes where it is either scales as in (3) or is a negative constant. In \cite{116}\cite{117}, it is argued that the general form of the scaling for systems with finite Fermi surfaces is for a leading term proportional to the product of the area between the subsystems and a logarithmic correction: a rescaling of the general picture on to the finite Fermi surface. Our expectation of the numerics from these results is therefore threefold: dependent on whether the lattice ensembles are gapped \cite{112}, gapless and conformally invariant \cite{euler} or gapless with finite Fermi surfaces \cite{116}\cite{117}. 

In Figure 3 the logarithm of the zeroes density in (19) is plotted as a function of the angle subtended along the unit circle in the complex-$(\beta J)^2$ plane, for an ensemble generated at $\Delta=2.0$. The scaling of the zeroes density can be fitted via two separate curves; where the density of the zeros sharply increases at the points $\phi\sim 0$ and $\phi\sim\phi'$, and also for the flat $\phi$-independent plateau. The scaling exponent values for fits to the second order scaling ansatz in (21) at $\phi\sim 0$ and $\phi\sim\phi'$ are tabulated in Table 1, along with the value of the intercept of the flat plateau, ${\rm{ln}}\Delta\tau_{IR}$.  
The fitted exponent values are different for ensembles generated at $\Delta=1.78$ (where $\tilde{c}\sim 1$) and $\Delta=2.0$ (where $\tilde{c}\sim 1/9$), whilst for $\Delta=1.01$ the measured scaling is purely first order. 
Although there is variation in the exponent values within the (jackknife) error estimates, it is important to note that the lattice units used to define $\mathbb{C}^{2}$ in the analysis are not held fixed. The comparison of finite size effects through volumes is therefore best made through ${\rm{ln}}\Delta\tau_{IR}$ rather than $L^{2}$. If the scaling exponent values are compared directly with comparable DMRG measurements the values show a similar level of consistency as a function of lattice volume \cite{st3-dm}\cite{st1-dm}. 

For ensembles generated in the vicinity of the N\'eel point, at $\Delta=1.01$, the $\mathbb{C}^{2}$ subsystem is analytically continued on to the boundary of the system $\theta_{1}\otimes\theta_{2}$. The flat plateau in the zeroes density indicates that the entanglement entropy scales linearly in the boundary area ($\theta_{1}\otimes\theta_{2}$) with a non-universal prefactor. There is no multiplicative logartihmic correction, and the scaling is therefore of the general form expected for conformally invariant critical points \cite{euler}\cite{112}. Any subleading universal correction is lost in the analysis, because the analytic continuation is not defined beyond the system boundary. 

This scaling picture differs  for the ensembles generated at $\Delta=1.78$ and $\Delta=2.0$, where the $\mathbb{C}^{2}$ subsystem is analytically continued on to the boundary of the subsystem $\phi_{1}\otimes\phi_{2}$, and not on to the boundary of the system $\theta_{1}\otimes\theta_{2}$. For ensembles generated in the vicinity of the first order transition between the Ising phase and quantum phase above the N\'eel point, at $\Delta=1.78$ \cite{11}\cite{138}, the flat plateau in the zeroes density indicates that the entanglement entropy scales linearly in the boundary area of the subsystem ($\phi_{1}\otimes\phi_{2}$) with a non-universal prefactor and also logarithmically with a universal prefactor $\tilde{c}$. However, the $\phi_{1}\otimes\phi_{2}$ subsystem itself also scales linearly on to $\theta_{1}\otimes\theta_{2}$ from the operator definition in (4) (this corresponds to the global mismatch between $\phi_{1}\otimes\phi_{2}$ and $\theta_{1}\otimes\theta_{2}$). Hence the scaling follows the general form for gapless systems with a finite Fermi surface in \cite{116}\cite{117}, of area law times a logarithmic correction. 

Even though the mismatch between $\phi_{1}\otimes\phi_{2}$ and $\theta_{1}\otimes\theta_{2}$ is a non-universal factor, by constructing exact operators for $\mathbb{C}^{2}$ we are able to extract the central charge and Euler characteristic for the finite Fermi surfaces of these ensembles. At $\Delta=1.78$ the ensembles lie in the vicinity of the first order transition between the Ising phase and quantum phase above the N\'eel point, whereas at $\Delta=2.0$ the ensembles are in the Ising regime \cite{11}\cite{138}. 
In both cases the central charge for the primary chiral field in $\mathbb{C}^{2}$ is $c=\overline{c}=1/2$, but the Euler characteristics differ as the later case the $Z_{2}$ symmetry is broken. The Euler characteristics for the surfaces of the ensembles are defined by counting the number of vertices of the hypercubes in (8), subtracting the number of edges and adding the number of faces. At $\Delta=1.78$ the Euler characteristic is given by $12/3 -8/2 +6 = 6$, whereas at $\Delta=2.0$ the Euler characteristics is given by.
$8/3 -8/2 +2 = 2/3$. From (3), the universal prefactors of the scaling of the boundary $\phi_{1}\otimes\phi_{2}$ should therefore be $\tilde{c}\sim 1$ and $\tilde{c}\sim 1/9$, which are consistent with the values in Table 1.

\begin{figure}
\epsfxsize=4.5 in\centerline{\epsffile{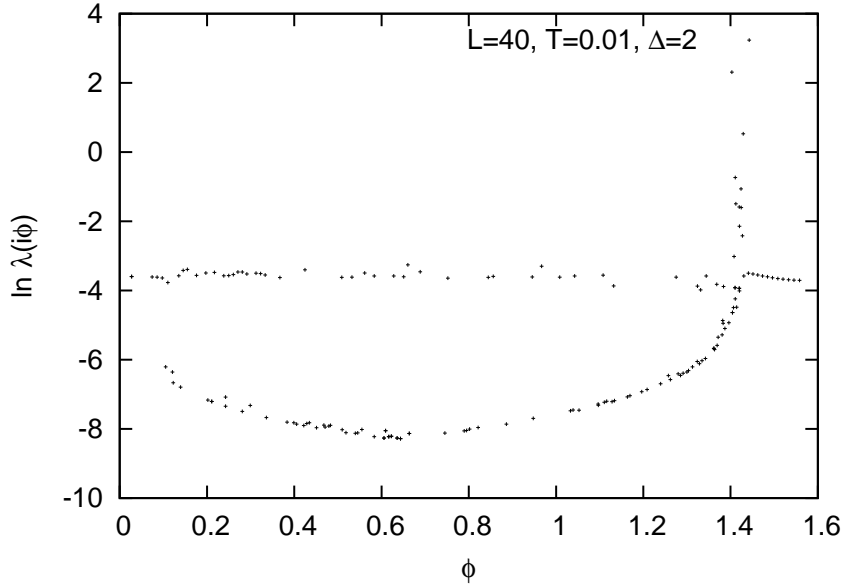}}
\caption{The logarithm of the zeroes density ${\rm{ln}} \, \Lambda(\phi)$ $(={\rm{ln}} \, \lambda(i\phi))$ of the spin-1/2 antiferromagnet on the square lattice plotted as a function of the angle $\phi$ subtended from the real axis of the complex-$K^{2}$ plane to a given zero along the unit circle. The lattice size of this ensemble is $L^2 = 40\times 40$, the inverse temperature is $\beta=100$, and the XXZ anisotropy parameter is $\Delta=2.0$. }
\end{figure}

\begin{table}
\begin{center}
\begin{tabular}{|l|l||r|r|r|r|r|}
\hline
$L$        & $\Delta$      & $\tilde{c}_{\,\,\phi \, \sim \, 0\,\,}$  	& $\tilde{c}_{\,\,\phi \,\sim\, \phi'\,\,}$ 	 & ${\rm{ln}}\Delta\tau_{IR}$ \\
\hline
\hline
12	   & 1.78	   & 1.0332(0.1903) 		& 1.0781(0.0182) 	&-2.3983(0.0090)\\
\hline
16	   & 1.78	   & 1.0177(0.0226) 		& 1.0160(0.0256) 	&-2.4309(0.0025)\\
\hline
20	   & 1.78	   & 1.0097(0.0189) 		& 1.0074(0.0648)	&-2.4291(0.0021)\\
\hline
24	   & 1.78	   & 1.0073(0.0250) 		& 0.9999(0.2222) 	&-2.4236(0.0017) \\
\hline
32	   & 1.78	   & 1.0101(0.0065)       	& 0.9997(0.0167) 	&-2.4245(0.0012) \\
\hline
40	   & 1.78	   & 1.0066(0.0480)		    & 1.0000(0.0119)	&-2.4052(0.0003) \\
\hline
\hline
$L$        & $\Delta$      & $\tilde{c}_{\phi \sim 0}$  	& $\tilde{c}_{\phi \sim \phi'}$ 	& ${\rm{ln}}\Delta\tau_{IR}$ \\
\hline
\hline
12	   & 2.00	   &  0.1102(0.0952) 		&  0.1092(0.0526) 	& -2.5679(0.0132) \\
\hline
16	   & 2.00	   &  0.1502(0.0505)    	&  0.1547(0.0045) 	& -2.3807(0.0022)\\
\hline
20	   & 2.00	   &  0.1504(0.0471) 		&  0.1305(0.0034) 	& -2.4685(0.0017) \\
\hline
24	   & 2.00	   &  0.1401(0.0057) 		&  0.1076(0.0058) 	& -2.4863(0.0010)\\
\hline
32	   & 2.00	   &  0.1337(0.0067) 		&  0.1077(0.0065) 	& -2.4690(0.0002)\\
\hline
40	   & 2.00	   &  0.1226(0.0476) 		&  0.1172(0.0313) 	& -3.5141(0.0038) \\
\hline
\end{tabular}
\end{center}
\caption{Dependence of the scaling exponent $\tilde{c}$ (determined from the fit of the logarithm of the zeros density to (\ref{asymp})) on the lattice system length $L$ and anisotropy $\Delta$, at a fixed inverse temperature of $\beta=100$. In the final column we give the non-analytic first order contribution to scaling, ${\rm{ln}}\Delta\tau_{IR}$, corresponding to the intercept of the flat plateau in Figure 2.}
\end{table}

\section{summary}

We have presented a new procedure in this article, which uses the convergence properties of an exact expansion of the lattice partition function of the continuous-time Quantum Monte Carlo method to identify the critical scaling of the spin-1/2 antiferromagnet on the square lattice for ensembles we have generated at fixed XXZ anisotropy values of $\Delta=1.01, \Delta=1.78$ and $\Delta=2.0$. This procedure is closely related to recent exact diagonalization studies of the scaling of the entanglement entropy of several quantum chain systems, and shows a similar level of accuracy \cite{st1-dm}\cite{st3-dm}\cite{st2}. The entanglement entropy in our approach is found by calculating the density of zeroes of a polynomial ring, and we have used this ring to represent the Chern-Simons phase difference between the two non-orthogonal polynomial representations that define the ground state of the spin-1/2 antiferromagnet on the square lattice in \cite{p937_1}. We have shown that the meromorphic convergence of the density of these zeroes arises from treating the replica method for entangelment entropy \cite{122} via a prescription for $\zeta$-function renormalization. We have compared our results with three different scaling pictures for the entangelement entropy of two dimensional fermionic quantum systems which are either gapped \cite{112}, gapless and conformally invariant \cite{euler} or are gapless with finite Fermi surfaces \cite{116}\cite{117}. Our approach allows us to identify the universality of the scaling associated with the primary chiral field of the spin-1/2 antiferromagnet on the square lattice, even though the Fermi surfaces of our ensembles may be finite and the general form of the scaling of the entanglement entropy in this case comes with a non-universal prefactor \cite{116}\cite{117}. This scheme is therefore useful for directly investigating the entanglement entropy of spin impurities in the underdoped cuprates that can be treated via the spin-1/2 antiferromagnet on the square lattice \cite{exp1}\cite{exp2}. Nothing in our construction forces the zeroes density we have obtained from numerics to scale in a meaningful way. However, in each of the ensembles we have generated, the polynomial zeroes show good evidence of an analytic continuation symmetry which validates the idea presented in \cite{200}\cite{p7215_1}\cite{149} that there is an intrinsic Chern-Simons term in the quantum regime of the spin-1/2 antiferromagnet on the square lattice that leads to the statistical transmutation of fermionic statistics from the UV into the IR. In the vicinity of the N\'eel point at $\Delta=1.01$, we have found that the scaling follows the form for gapless and conformally invariant critical points in 2d quantum systems \cite{euler}. This supports the general idea presented in deconfined quantum criticality \cite{17}\cite{GetPDF5}, that the N\'eel point can be smoothly deformed by quantum fluctuations to give a second order fixed point somewhere above $\Delta = 1$. For our ensembles in the Ising phase at $\Delta=1.78$ and $\Delta=2.0$ we have we have found the scaling follows the form for gapless with finite Fermi surfaces in 2d quantum systems \cite{116}\cite{117}. The Euler characteristics we have identified for the prefactor of the scaling associated with the primary chiral field of these ensembles are consistent with topology of the hypercubes we have used to construct the lattice operators in these two regions which are above and at the boundary for the Ising phase \cite{11}\cite{138}.

\end{document}